**Combining oligo pools and Golden Gate cloning to create protein variant libraries or guide RNA libraries for CRISPR applications.**

Alicia Maciá Valero[1#], Rianne C. Prins[1#], Thijs de Vroet[1#], Sonja Billerbeck [1*]

[1]Molecular Microbiology, Groningen Biomolecular Sciences and Biotechnology Institute, University of Groningen, Groningen, 9747 AG, The Netherlands

***Correspondence**: s.k.billerbeck@rug.nl

#These authors contributed equally. The authors have the right to change the order of their names appearing on the author list.

**Abstract:** Oligo pools are array-synthesized, user-defined mixtures of single-stranded oligonucleotides that can be used as a source of synthetic DNA for library cloning. While currently offering the most affordable source of synthetic DNA, oligo pools also come with limitations such as a maximum synthesis length (approximately 350 bases), a higher error rate compared to alternative synthesis methods, and the presence of truncated molecules in the pool due to incomplete synthesis. Here, we provide users with a comprehensive protocol that details how oligo pools can be used in combination with Golden Gate cloning to create user-defined protein mutant libraries, as well as single guide RNA libraries for CRISPR applications. Our methods are optimized to work within the Yeast Toolkit Golden Gate scheme, but are in principle compatible with any other Golden Gate-based modular cloning toolkit and extendable to other restriction enzyme-based cloning methods beyond Golden Gate. Our methods yield high-quality, affordable, in-house variant libraries.

**Keywords**
Golden Gate cloning, oligo pools, mutational libraries, CRISPR libraries

The **Supplementary Material** supporting this article is attached at the end of this document.

## 1. Introduction

Constructing diverse libraries of a sequence of interest is a cornerstone of biological engineering, enabling the creation and examination of multiple variants of a biological function in one pot.

Protein variant libraries, for example, can be used in protein engineering to perform mutational scans, where each amino acid is systematically exchanged for several or all other 19 standard amino acids.[1] The resulting deep mutational scanning (DMS) libraries can subsequently be functionally characterized using various methods as outlined in available reviews,[1, 2] allowing a protein scientist to create detailed sequence-function maps of a protein. Such maps have emerged as valuable tools to understand and engineer the substrate-specificity, solubility, and kinetics of enzymes,[3–5] the allosteric behavior of proteins,[6, 7] the clinical properties of therapeutic antibodies,[8] or the attachment affinity of viruses to their host's receptor based on mutated capsid proteins.[9, 10] DMS data even allows us to predict protein structures by using experimentally determined epistatic interactions between pairs of residues and the fact that epistasis correlates with structural proximity.[11]

Other types of mutational libraries can facilitate insertion or deletion scans, in which additional residues (*e.g.,* encoding restriction sites or affinity tags) are systematically inserted after each amino acid position in a protein. These libraries can subsequently be harnessed to identify permissive sites for tags[12] or to create intermediate libraries for protein minimization strategies.[13]

In addition, libraries for CRISPR applications often encompass a large number of different 20-base pair spacer regions that target specific DNA stretches in a host genome. These libraries can be combined with Cas9 and other Cas variants for CRISPRko (knockout), CRISPRi (inhibition), and CRISPRa (activation) screens. Such CRISPRx screens have been widely applied for functional genomics,[14–19] and gene circuit design [20, 21] across bacteria, yeast, and mammalian cells. For instance, this has enabled the identification of essential gene functions,[22] disease-related phenotypes, drug resistance profiles, and modes of action of drugs.[23]

A key requirement for creating such libraries is access to synthetic DNA that encodes the numerous defined mutational variants. Although the cost of double-stranded DNA has significantly dropped by several orders of magnitude over the last decades, the substantial number of base pairs required for constructing comprehensive mutational libraries continues to pose a bottleneck in terms of cost.

Purchasing pools of single-stranded oligonucleotides is currently the most affordable option for obtaining synthetic DNA, with a cost per base approximately 100-fold lower than that of double-stranded DNA (<0.001 compared to 0.05-1 USD per base).[24]

However, oligo pools come with certain limitations, including a maximum length constraint (currently 350 bases), a higher error rate compared to alternative gene synthesis methods, and the presence of truncated variants of the oligos due to incomplete synthesis.[24]

Here, we outline two methods that exemplify how oligo pools and Golden Gate cloning can be combined to create high-quality libraries for protein mutational scans or single guide RNA (sgRNA) libraries. We provide step-by-step methods for the design and creation of the libraries and provide recommendations for downstream screening and Next-Generation Sequencing (NGS)-based analysis.

Our protocol overcomes the aforementioned limitations by working with a gene-tiling approach. This approach enables the use of relatively short oligos with an acceptably low error rate. Furthermore, we use a double-stranding protocol that minimizes complications caused by truncated oligos.

The first method **(Method 1)** describes a protocol for cloning mutational protein libraries, exemplified with a mutational scan, where each residue is replaced by a selected set of amino acids, and an insertional scan, where an epitope tag is introduced at each position within a given protein. Both examples cover the length of a full open reading frame (ORF), but the methods can also be used for a more focused mutagenesis of specific regions within an ORF. The second method **(Method 2)** describes the cloning of a single guide RNA (sgRNA) library for CRISPRi applications.

The methods outlined herein make use of the genetic components from the Yeast Toolkit (YTK)[25], which has been developed for the yeast *Saccharomyces cerevisiae* and has been extended to other yeast such as *Pichia pastoris*[26] and *Candida glabrata*.[27] However, the methods should be rapidly adjustable to other Golden Gate-based toolkits for bacteria,[28][29] plants[30] or mammalian cells.[31]

We further note that, while we limit ourselves here to a few selected examples, the presented methods can be extended to any other Golden Gate scheme or even traditional restriction-enzyme-based cloning methods.

## 2. Materials

The execution of the described protocols requires the following standard laboratory equipment and consumables. While certain providers are mentioned in the manuscript, the materials described may be obtained from any other similar suppliers.

**1.** Standard reagents, consumables and instrumentation for thermocycler reactions.
**2.** Standard reagents and equipment for agarose gel electrophoresis.
**3.** Standard reagents, consumables and instrumentation for microbial culturing.
**4.** Standard reagents, consumables and instrumentation for transformation of *Escherichia coli* and *S. cerevisiae*.
**5.** Standard reagents, consumables and instrumentation for the purification of PCR products, extraction of DNA from agarose gels and plasmid extraction.

**6.** Access to Sanger sequencing and NGS services.
**7.** Access to a Transilluminator or transillumination goggles and a UV lamp.

## 2.1. Plasmids

All plasmids relevant to this protocol are listed in **Table 1**. These plasmids are either available via the YTK on Addgene **(**MoClo-YTK Plasmid Kit #1000000061) or upon reasonable request from the authors. The plasmids are organized by Method and some mentions are redundant but necessary to follow the workflow. Several plasmids mentioned herein were only created *in silico* to exemplify the workflow and the links to the full sequences for those *in silico* plasmids can be found in **Supplementary Table 1.** Notably, the protocol described below can be adjusted to other (type IIS) restriction enzymes, overhang sequences and antibiotic resistance markers.

**Table 1. List of plasmids.**

| Name | Specification | Comment | Availability/Ref |
|---|---|---|---|
| **Method 1** | | | |
| pYTK002 | ConLS, type 1 | Used to assemble pRS413-pTEF2-Venus-tENO1 | Addgene # 65109/ |
| pYTK014 | TEF2 promoter, type 2 | | Addgene # 65121/[25] |
| pYTK033 | Venus ORF, type 3 | | Addgene # 65140/[25] |
| pYTK051 | ENO1 terminator, type 4 | | Addgene # 65158/[25] |
| pYTK072 | ConRE, type 5 | | Addgene # 65179/[25] |
| pYTK076 | HIS3, type 6 | | Addgene # 65183/[25] |
| pYTK081 | CEN6/ARS4, type 7 | | Addgene # 65188/[25] |
| pYTK083 | AmpR-ColE1, type 8 | | Addgene # 65190/[25] |
| pYTK047 | GFP-dropout, type 234 | Used as a template to clone of the destination vectors | Addgene # 65154/[25] |
| pRS413-pTEF2-Venus-tENO1 | pRS413-type vector containing the Venus gene under control of the TEF2 promoter and the ENO1 terminator. (assembled via Golden Gate using the YTK) | Used to exemplify the segmentation for Method 1 | Designed *in silico.* Links to sequences are given in **Supplementary Table 1** |
| Destination vector 1 | pRS413-pTEF2-Venus-tENO1 with **Segment 1** replaced by a GFP-dropout | | |
| Destination vector 2 | pRS413-pTEF2-Venus-tENO1 with **Segment 2** replaced by a GFP-dropout | | |
| Destination vector 3 | pRS413-pTEF2-Venus-tENO1 with **Segment 3** replaced by a GFP-dropout | | |
| Destination vector 4 | pRS413-pTEF2-Venus-tENO1 with **Segment 4** replaced by a GFP-dropout | | |
| Destination vector 5 | pRS413-pTEF2-Venus-tENO1 with **Segment 5** replaced by a GFP-dropout | | |
| Destination vector 6 | pRS413-pTEF2-Venus-tENO1 with **Segment 6** replaced by a GFP-dropout | | |
| Destination vector 7 | pRS413-pTEF2-Venus-tENO1 with **Segment 7** replaced by a GFP-dropout | | |
| Destination vector 8 | pRS413-pTEF2-Venus-tENO1 with **Segment 8** replaced by a GFP-dropout | | |
| Destination vector 9 | pRS413-pTEF2-Venus-tENO1 with **Segment 9** replaced by a GFP-dropout | | |
| Destination vector 10 | pRS413-pTEF2-Venus-tENO1 with **Segment 10** replaced by a GFP-dropout | | |
| **Method 2** | | | |
| pCgTK01 | CgCEN/ARS, type 7 | Used to assemble pRS414-type vector via Golden Gate | From authors; [27] |
| pCgTK02 | TRP1, type 6 | | From authors; [27] |
| pYTK002 | ConLS, type 1 | | Addgene # 65109/[25] |
| pYTK047 | GFP-dropout, type 234 | | Addgene # 65154/[25] |
| pYTK072 | ConRE, type 5 | | Addgene # 65179/[25] |
| pYTK084 | KanR-ColE1, type 8 | | Addgene # 65191/[25] |
| pYTK50 | sgRNA dropout, type 234 | Used as a template to clone the gRNA expression cassette into the above pRS414-type vector, resulting in the gRNA Destination vector | Addgene # #65157/[25] |
| gRNA Destination vector | pRS414 with gRNA expression cassette based on pYTK050. 20 bp spacer replaced by a GFP dropout. | Used for cloning the gRNA library | Unpublished. Link to sequence in **Supplementary Table 1** |

### 2.2. Single-stranded oligonucleotides and oligo pools

Oligonucleotides used in this protocol are provided in **Table 2.** 100 µM stocks were generated by resuspending the lyophilized oligonucleotides in the appropriate amount of $ddH_2O$ and subsequently stored at -20 °C, from which 10 µM working stocks were generated using $ddH_2O$. Examples of *in silico* designed oligo pools are provided in **Supplementary Tables 2 to 4).** 10 µM stocks of oligo pools were generated in $ddH_2O$ and stored at -20 °C.

**Table 2. List of primers.**

| # | Name | Sequence 5'-3' | Comment |
|---|---|---|---|
| **Method 1** | | | |
| 1 | segment 2 - 5'flank fw | GCAT **CGTCTCA** TATG tctaaaggtgaagaat | *In silico* designed primers suggested to clone Destination Vector 2. Bolt-capital letters: BsmBI restriction site. Underlined-capital letters: specific 4-bp overhangs. Lowercase letters: sequence for priming on the template. |
| 2 | segment 2 - 5'flank rv | GCAT **CGTCTCA** ACCA ttaacatcaccatc | |
| 3 | segment 2 - 3'flank fw | GCAT **CGTCTCA** ACTG gtaaattgccag | |
| 4 | segment 2 - 3'flank rv | GCAT **CGTCTCA** GGAT ttatttgtacaattcatcc | |
| 5 | GFP dropout fw | GCAT **CGTCTCA** TGGT tgagaccgaaag | |
| 6 | GFP dropout rv | GCAT **CGTCTCA** CAGT tgagacctataaac | |
| 7 | oPool reverse | TGCCGTCTCAGGTCTCA | Reverse primer used for double-stranding the oligo pool in Method 1, binds to oligo pool-encoded landing pad (**Supplementary Tables 3-5** exemplify oligo pool sequences). |
| 8 | seq seg 2 fw (with partial Illumina adapter) | ACACTCTTTCCCTACACGACGCTCTTCCGATCT ctataattaactaaacagatct | Primers for amplification of Segment 2 for NGS. The amplicon size is 260 bp which is a suitable amplicon size 250 bases paired-end reads. Capital letters: Partial Illumina adapter. Lowercase letters: sequence for priming on the Destination vector 2. No barcodes for multiplexing have been added. |
| 9 | seq seg 2 rv (with partial Illumina adapter) | GACTGGAGTTCAGACGTGTGCTCTTCCGATCT gtttcatatgatctgggta | |
| **Method 2** | | | |
| 10 | sgRNA oPool reverse | CTGCCGTCTCAAAAC | Reverse primer used for double-stranding the oligo pool in Method 2. |
| 11 | YTK050_BsaI_fw | CTAG **GGTCTCG** TGCT **tatccactagacagaagtttgcgtt** | *In silico* designed primers suggested to amplify sgRNA acceptor from YTK050. Bolt-capital letters: BsmBI restriction site. Underlined-capital letters: specific 4-bp overhangs. Lowercase letters: sequence for priming on the template. |
| 12 | YTK050_BsaI_rv | CTAG **GGTCTCG** CTGA **atgtgcttcagtattacatttttgcct** | |
| 13 | Illumina sequencing_adapter FWD2 | AGACGTGTGCTCTTCCGATCTaaaacttcggtcaagtcatct | Primers for amplification of the 20 bp spacer region for NGS. Capital letters: Partial Illumina adapter. Lowercase letters: for priming on the gRNA Destination vector. |
| 14 | Illumina sequencing_adapter REV2 | CTACACGACGCTCTTCCGATCTgataacggactagccttatttt | |

### 2.3. Enzymes and buffers

**1.** T7 Ligase
**2.** BsaI-HFv2
**3**. BsmBI-v2
**4.** 2x Phire Hot Start II PCR Master Mix
**5.** 2x Phusion High-Fidelity PCR Master Mix
**6.** 10x NEB Buffer 3
**7.** 10x T4 ligase buffer

### 2.4. Media, antibiotics, and other consumables

**1.** Ampicillin (100 mg/mL stock), and Kanamycin (50 mg/mL stock) dissolved in sterile $H_2O$ and stored in 1 mL aliquots at -20 °C. The stock concentration is 1000x.
**2.** *E. coli* growth medium: LB liquid and solid media: tryptone 10 g/L, yeast extract 5 g/L, NaCl, 5 g/L, add 20 g/L agar for solid media.
**3.** Yeast growth medium: YPD liquid and solid media: yeast extract 10 g/L, peptone 20 g/L, dextrose 20 g/L, add 20 g/L agar for solid media; Synthetic complete (SC) dropout liquid and solid media: dextrose 20 g/L, yeast nitrogen base 6.7 g/L, 50 mL/L amino acid dropout solution (20x, **Table 3**) supplemented with or without the auxotrophic selection components: 2 g/L uracil, 10 g/L tryptophan, 10 g/L leucine and 10 g/L histidine, add 20 g/L agar for solid media.
**4.** Yeast plate mixture (10 mL total): Mix 8.9 mL sterile-filtered 50 % PEG 4000, 1 mL 1 M lithium acetate, 100 µL 1 M Tris-HCl (pH 7.5), 20 µL 0.5 M EDTA-NaOH (pH 8.0) and vortex.
**5.** PCR reaction tubes
**6.** Eppendorf tubes
**7.** Standard Petri dishes

### 2.5. Strains

For cloning and plasmid maintenance, we use *E. coli* DH5α; for transformation of plasmids into yeast, we use *S. cerevisiae* BY4741.[32]

## 3. Methods

### 3.1. Method 1. Creating protein variant libraries using oligo pools and Golden Gate cloning.

We provide protocols for two types of protein variant libraries: 1) an insertional scan, where an epitope tag is inserted after every residue in a protein of interest and 2) a mutational scan where each amino acid is substituted with one of six other selected amino acid residues. The general workflow is illustrated in **Figure 1.**

Two important considerations should be kept in mind when using oligo pools for cloning such libraries: the length limitation of oligo synthesis and the single-stranded character of the oligonucleotides obtained.

Firstly, the maximum oligo length possible for synthesis (350 bases) may often be shorter than the gene of interest. Consequently, our method involves tiling a gene into "segments" that are short enough to be covered by oligos **(Figure 2A).** In fact, to reduce synthesis errors we keep them below 100 bases, as the chance for error increases with size. Subsequently, we prepare a destination vector for each segment, which can accept the corresponding oligos from the oligo pool and serve to clone each segment library **(Figure 2A and B).** Eventually, all individual segment libraries can be pooled into a single final library. However, keeping individual segment libraries separate offers advantages in terms of quality control and troubleshooting per segment. For instance, if NGS-based sequencing reveals that a particular segment is incomplete or contains off-target mutations, the segment can be rapidly re-cloned. Moreover, the destination vectors are reusable for other purposes and this approach allows a swift creation of focused libraries in specific regions of a protein, *e.g.* as a follow-up after an initial round of full-gene mutagenesis.

The second consideration concerns the single-stranded nature of oligonucleotides, necessitating their conversion to a double-stranded form before use. Instead of using a conventional PCR approach for the double-stranding and amplification of the oligo pool, we have identified a crucial step: The double-stranding of the oligos through a single primer extension reaction using a reverse primer **(Figure 3).** This is crucial because not all oligos in a pool attain full-length synthesis due to technical limitations in the synthesis process **(Figure 3A).** Consequently, an oligo-pool contains truncated oligos, which may or may not encode an intended mutation. Given the high sequence similarity among oligos within an oligo pool designed for mutational scanning libraries, with variations mainly confined to the introduced tag or mutated position, these truncated versions can function as primers when using multiple PCR cycles, leading to crossovers between different oligos within the oligo pool. This, in our own experience, results in libraries that contain noise in the form of double-stranded oligos harboring non-designed mutations or even wild-type sequences **(Figure 3B).** A single reverse primer extension ensures that only full-length oligos within the pool become double-stranded, and are able to participate in the subsequent reactions.

We organized Method 1 into two steps (**Figure 1)**:

**3.1.1 Design of the library**: this includes, 1. segmentation of the gene of interest; 2. design of the destination vectors based on the segmentation and 3. design of the oligo pools.

**3.1.2. Creation of the library**: this includes, 1. cloning of the destination vectors; 2. double-stranding the oligos; 3. performing the Golden Gate reaction; 4. transformation of *E. coli* with the mutational library; 5. evaluation of the library coverage and considerations for oversampling; 6. pooling the library; 7. transformation of the library into yeast (*S. cerevisiae*). Further, we provide suggestions for selection and/or screening assays (8.); and for library quality control and genotype-phenotype linking via NGS (9.).

All plasmid designs are provided as links to the gene editing software *Benchling* (www.Benchling.com), but any gene editing software is suitable.

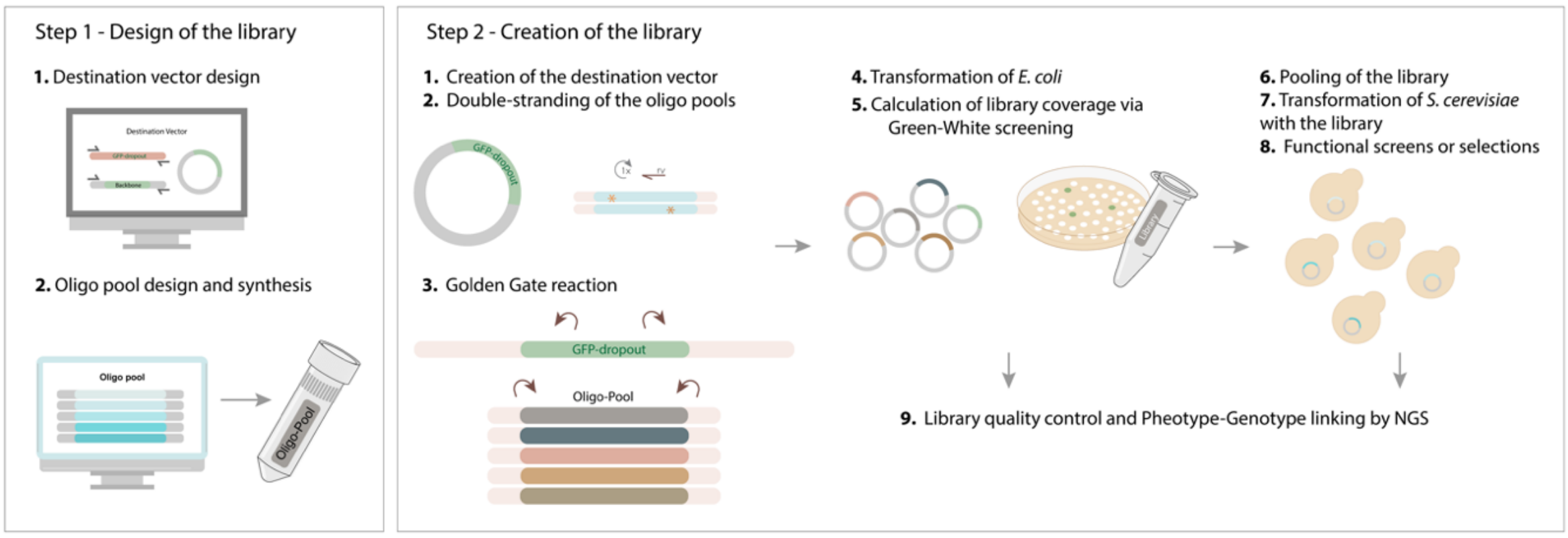

**Figure 1. Overview of the library creation workflow using oligo pools.** We have divided the method into two steps: Step 1 involves the *in-silico* design of the library and step 2 describes the actual creation of the library. The numbered sub-steps refer to the sub-steps outlined in this protocol.

**3.1.1. Design of the library.**

Here, we use the ORF encoding the fluorescent Venus protein (from pYTK033) as an example for the two full-length protein mutagenesis approaches. The gene was cloned *in silico* into a pRS413-type vector flanked by a TEF2 promoter and an ENO1 terminator using YTK parts **(Table 1).** For cloning a user-defined gene into the YTK format, follow the steps outlined in M. E. Lee *et al.,* (2015).[25]

**1. *In silico* segmentation of the gene of interest.** The *Venus* gene (over 700 bp) needs to be divided into segments (‘tiles’) that can be covered by the oligos within an oligo pool. Parameters that need to be weighed are: First, the length of the segments; second, the number of segments; and third, the sequence of the 4 bp overhangs that are generated by restriction digestion and used for ligation (**Figure 2**). The oligos should not be too long as the number of full-length molecules in a pool decreases and the probability of synthesis errors increases when increasing oligo synthesis length **(Figure 3A).**
On the other hand, the workload associated with cloning individual destination vectors increases with an increasing number of segments. Taking these factors into account, we have had a positive experience working with segments of 15 amino acids (45 bp), 16 amino acids (48 bp) and 27 amino acids (81 bp) in length.

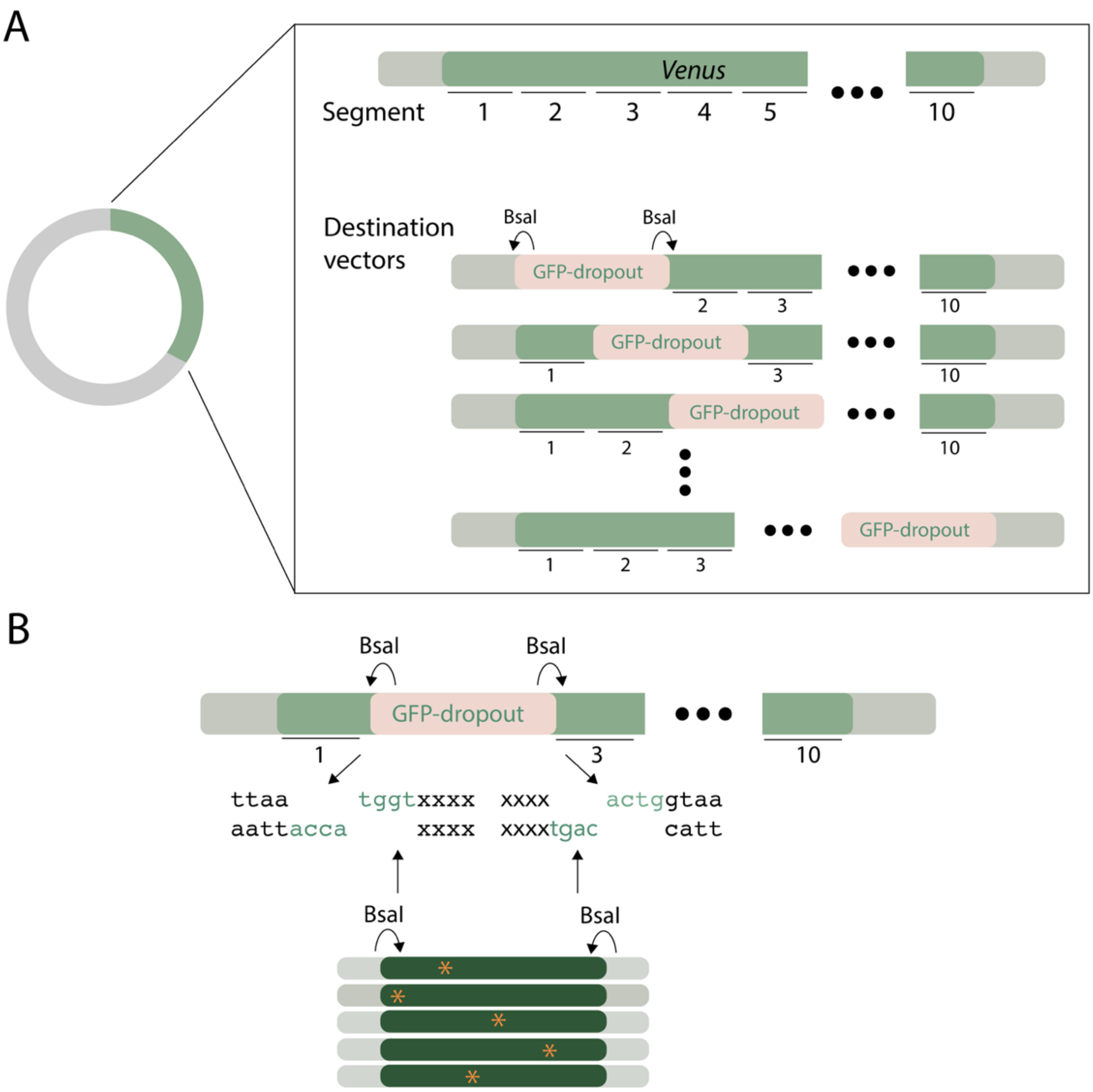

**Figure 2. Segmentation of a gene and cloning of the destination vectors. A.** The gene is divided into segments. For each segment, a destination vector is cloned by replacing the segment with a GFP-dropout cassette. **B.** The GFP-dropout cassette present in every destination vector already encodes outward-cutting BsaI sites that generate 4 bp overhangs. The overhangs of the destination vector are complementary to the overhangs similarly generated in the oligo pool of the respective segment library and are used to assemble the library constructs in a Golden Gate reaction.

Here, we divided the *Venus* gene into 10 segments, each spanning 23-27 amino acids. We adjusted the segment length based on the predicted ligation efficiency of the 4 bp overhangs. The 5' and 3' overhangs of a given segment are defined by the 4 bp flanking it at each side. Ligation efficiency for each overhang pair can be calculated using the NEB Fidelity Viewer® (https://ligasefidelity.neb.com/viewset/run.cgi). The resulting 4 bp overhangs for the Venus segmentation and their estimated ligation efficiencies are then as follows: Segment 1: ATCT and CACA (100 %); Segment 2: TGGT and ACTG (100 %); Segment 3: TACT and TACC (93 %); Segment 4: TAGA and AAAG (100 %); Segment 5: TTTC and GAAT (92 %); Segment 6: AATC and AATG (99 %); Segment 7: TCAC and GGTG (100 %); Segment 8:

AGAT and CATT (100 %); Segment 9: CAAC and TTGT (100 %); Segment 10: GGTC and ATCC (100 %).

Another secondary factor that should be considered is that, for the creation of the destination vectors, one will need to amplify the 3' and 5' regions flanking the segment **(Supplementary Figure 1A).** As such, a functional primer binding site at these specific regions is also preferred (see section 3.1.2 1. Cloning of the destination vectors).

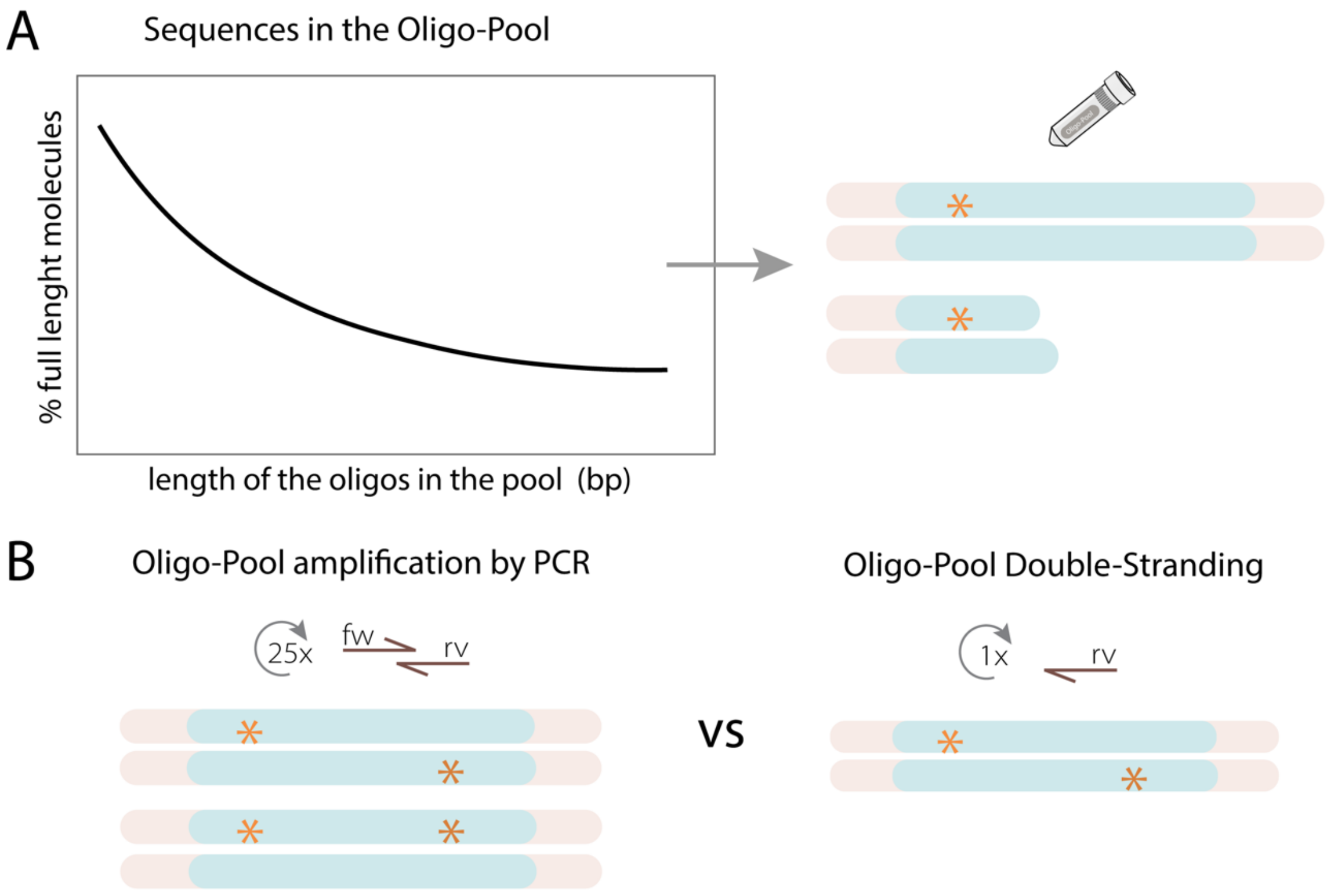

**Figure 3. Oligo-pool amplification versus double-stranding. A. Left panel:** Sequences in an oligo-pool are not all full-length due to current technical limitations in synthesis. As such, a mutagenesis oligo pool contains truncated oligos that may or may not encode an intended mutation. **Right panel:** Truncated oligos are present and those containing similar sequences may act as primers on complementary full-length oligos in a PCR reaction. **B.** Comparison between oligo pool amplification by PCR and one-step extension with a single reverse primer. Amplification by PCR (left) can cause final products that may contain unintended combinations of designed mutations, or even wild-type sequences. Oligo pool double-stranding (right) allows only the full-length synthesized oligos to get double-stranded and utilized in the subsequent Golden Gate reaction.

**2. *In silico* design of a destination vector for each segment.** Given the 10 segments, 10 destination vectors were designed. For each destination vector, the corresponding segment in the original gene was replaced with a GFP-dropout cassette. The GFP-dropout cassette can be expressed in *E. coli* and allows easy assessment of the Golden Gate efficiency via Green-White screening after transformation **(3.1.2. step 5.)**. We used the GFP-dropout cassette encoded on the YTK-vector pYTK047 **(Table 1)**, which includes outward cutting BsaI recognition sites. When replacing a gene segment, these sites generate specific overhangs in the original gene sequence, enabling subsequent seamless ligation of oligo pool sequences into the destination vector. The detailed sequences of the destination vectors can be found in **Supplementary Table 1** and the concept is illustrated in **Figure 2.** Note that, when using a cloning system other than YTK, precautions should be taken to ensure the absence of additional BsaI restriction sites in both the vector and the dropout cassette.

**3. Design of the oligo pools for each segment.** We will outline the design with two examples:

Example 1: Insertional scanning library.

We exemplify the oligo pool design for an insertional scanning library with the insertion of an ALFA-tag after every amino acid position of segment 2 of the Venus protein. The ALFA-tag is a 15-residue epitope tag (PSRLEEELRRRLTEP) that is recognized by an engineered nanobody that can be used for imaging, immunoprecipitation and protein purification.[33]

Follow the steps of the oligo pool design (**Figure 4**):

**1.** Reverse translate the amino acid sequence of the ALFA-tag using codons optimized for the expression host - in our case *S. cerevisiae* (resulting in 5'-TCTAGATTGGAAGAA GAATTGAGAAGAAGATTGACTGAA-3').

**2.** Insert this sequence in-frame after each codon of segment 2 *in silico*. Always verify that you did not create additional BsaI recognition sites at undesired positions.

**3**. Add the four segment-specific bases (four bases flanking the segment, yielding the overhangs after BsaI cleavage).

**4.** Add a BsaI-recognition site to the 5' and 3' ends of each oligo, making sure that the directionality of the cut is in the 3' direction.

**5.** At the 5' and 3' end of each oligo, add at least three extra nucleotides to improve the digestion efficiency. Historically, we used the sequence 5'-GTCTCTCATC-3'. This also yields a primer landing pad that will be used for double-stranding **(Figure 4C** and **Table 2)**. When using the same landing pad for all oligos, only one primer will be required for double-stranding an entire oligo pool. However, in case a user wants to encode multiple libraries in one oligo pool and selectively double-strand a sub-pool of oligos, different orthogonal landing pads can be used.

**6.** Encode each variant on a separate oligo to obtain equimolar amounts of each variant. Stoichiometric ratios between variants can be shifted by duplicating (or triplicating etc.) the number of certain oligos within the ordering list.

The complete oligo pool list for segment 2 is available in **Supplementary Table 2, (**see **Note 1** for calculating the number of nucleotides and estimating the expected price of the pool).

Example 2. Mutational scanning library. The design of a mutational scanning library follows the same principles. Here, we provide an example illustrating how every amino acid position of a protein can be mutated to either alanine (A), aspartic acid (D), serine (S), phenylalanine (F), leucine (L) or proline (P), which are representatives of amino acids with different physical-chemical properties. In case a position already encodes for one of those six amino acids, the pool list omits the oligo that would encode for that original wild-type sequence. The mutational scanning of the first codon of segment 2 is shown in **Table 4.** The full list of oligo pool sequences of segment 2 with the additional overhangs, BsaI restriction sites and nucleotides for better digestion efficiency (as outlined in **Figure 4B**) are provided in

**Supplementary Table 3.** This also generates a primer landing pad for double-stranding with the reverse primer (see **3.1.2. step 2.**). Note that in mutation S28L in **Supplementary Table 3** the codon *tta* was used instead of *ttg* in order to prevent the creation of additional BsaI recognition sites.

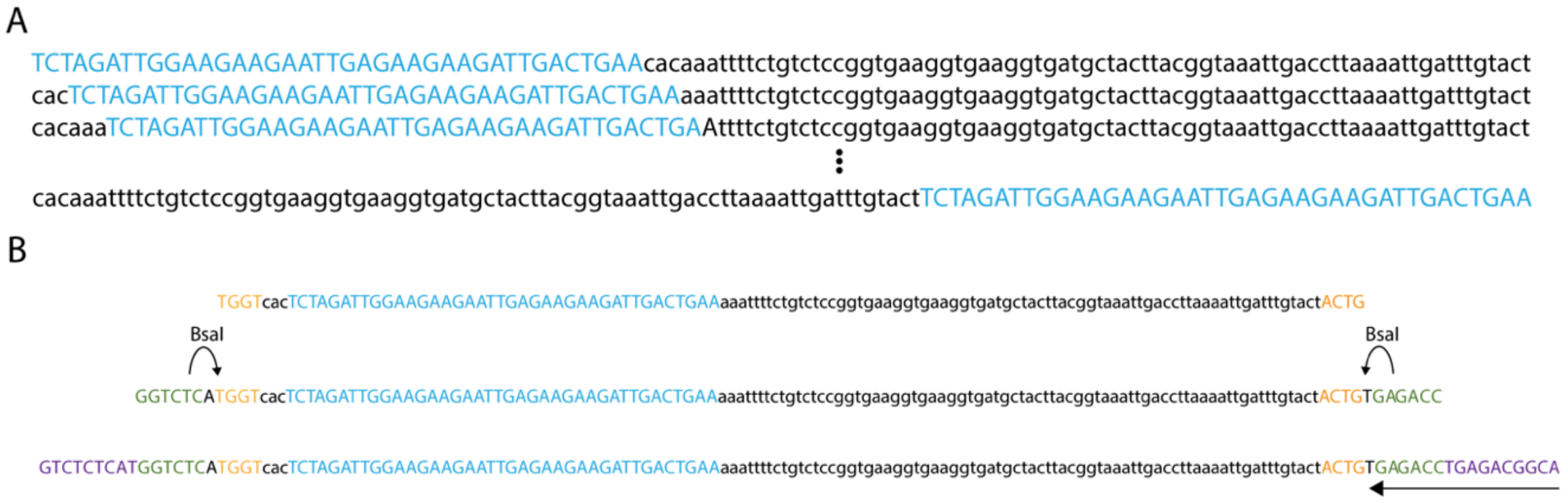

**Figure 4. Oligo pool design of an insertional scanning library. A:** The oligo pool is designed by inserting the desired tag (blue) after each codon in the segment. **B:** Additional nucleotides need to be added at the 5' and 3' end: First, the appropriate overhangs for the respective segment that allow specificity and directionality in ligation (orange); second, the BsaI-recognition sites in the correct orientation (green), plus one spacer nucleotide (black) as BsaI cuts one nucleotide from the recognition site in the 3' direction. Last, additional bases at the oligos' 5' and 3'ends that enhance cutting efficiency and generate a primer landing pad at the 3' end of each oligo for double-stranding (purple). The black arrow at the right bottom marks the landing pad sequence that can be used as reverse primer to double-strand all the oligos of the library.

Depending on the gene length, a mutational library can quickly expand, resulting in increased costs. If necessary, the usage of degenerate bases can mitigate the cost of a library. Degenerate or mixed bases are equimolar mixtures of two, three or four bases at a certain position within a sequence. **Supplementary Table 4** exemplifies the use of the mixed bases *m* (*c/a*), *k* (*g/t*) and *y* (*c/t*). Consequently, rather than using the separate codons *gct* and *gat* for adenine and aspartic acid, respectively, the unified codon *gmt* can encode both. The use of degenerate bases does not affect the total library size (amount of possible different variants), but the varying probabilities associated with each base should be considered during subsequent experimental phases to ensure that all variants are present.

**Table 4. Example of partial oligos designed for a mutational scan.** The first codon of segment 2 encodes for a histidine (H). This histidine was exchanged (in blue) for codons encoding either alanine (A), aspartic acid (D), serine (S), phenylalanine (F), leucine (L) or proline (P) using the codon with the highest usage frequency in *S. cerevisiae*.

| Amino acid | Codon mutation | Sequence |
|---|---|---|
| H | - | cacaaattttctgtctccggtgaaggtgaaggtgatgctacttacggtaaattgaccttaaaattgatttgtact |
| H→A | gct | **gct**aaattttctgtctccggtgaaggtgaaggtgatgctacttacggtaaattgaccttaaaattgatttgtact |
| H→D | gat | **gat**aaattttctgtctccggtgaaggtgaaggtgatgctacttacggtaaattgaccttaaaattgatttgtact |
| H→R | aga | **aga**aaattttctgtctccggtgaaggtgaaggtgatgctacttacggtaaattgaccttaaaattgatttgtact |
| H→S | tca | **tca**aaattttctgtctccggtgaaggtgaaggtgatgctacttacggtaaattgaccttaaaattgatttgtact |
| H→F | ttt | **ttt**aaattttctgtctccggtgaaggtgaaggtgatgctacttacggtaaattgaccttaaaattgatttgtact |
| H→L | ttg | **ttg**aaattttctgtctccggtgaaggtgaaggtgatgctacttacggtaaattgaccttaaaattgatttgtact |
| H→P | cca | **cca**aaattttctgtctccggtgaaggtgaaggtgatgctacttacggtaaattgaccttaaaattgatttgtact |

### 3.1.2. Creation of the library.

**1. Cloning of the destination vectors.** Here we provide recommendations and one example of how to clone the destination vectors. However, these can be cloned using any molecular cloning method of choice. The result should be 10 destination vectors in which each segment has been replaced by a GFP-dropout cassette. This cassette should contain 5' directed BsaI restriction sites, such that digestion of the entry vector leaves compatible overhangs within the original sequence for ligation of the corresponding oligo pool and results in an in-frame replacement of the wild-type segment. The GFP-dropout cassette (from pYTK047) and the original gene regions flanking the segment of interest can be separately amplified by PCR using overlapping primers or primers encoding for restriction enzymes and assembled into a vector by Gibson assembly, yeast assembly, fusion PCR or cloned via restriction digestion and ligation. In case of scanning mutagenesis of an entire gene, one can consider keeping the start- and stop-codon in place.

An example for creating the Destination Vector 2 via restriction-ligation cloning using the type IIS restriction enzymes BsaI and BsmBI is given below and illustrated in **Supplementary Figure 1A:**

**1.** Use primers 1 to 6 listed in **Table 2,** (see **Note 2**) to amplify the GFP-dropout cassette (template: pYTK047) and the 5' and 3' regions flanking Segment 2 (template: pYTK033).

**2.** Mix the YTK components required for assembly of the pRS413-type vector (pYTK002, pYTK072, pYTK076, pYTK081, pYTK083) and the TEF promoter (pYTK014) and ENO terminator (pYTK051) and digest with BsaI. Heat-inactivate BsaI after the digest.

**3.** Digest PCR-amplified segment 2-specific GFP dropout, and 5' and 3' flanking regions with BsmBI. Heat-inactivate BsmBI after the digest.

**4.** Perform a ligation reaction of all components, transform *E. coli* competent cells with the ligation mix and pick a few green colonies to verify the correct destination vector sequence. After sequence confirmation, this vector can be used for many applications of mutagenesis.

**2. Generation of double-stranded oligos.** Oligonucleotides in the pool are single-stranded. In order to generate double-stranded DNA, we run a single primer extension reaction using the reverse primer that binds to the landing pad (oPool reverse: 5'-TGCCGTCTCAGGTCTCA-3', **Table 2**, **Figure 4B**). Follow the steps:

**1.** Mix the following components in a PCR tube:

| **Primer extension mix** | |
|---|---|
| Reagent | Volume |
| Single-stranded oligos (1 µM) | 1 µL |
| Primer oPool_reverse (10 µM) | 2.5 µL |
| 2x Phire Hot Start II PCR Master Mix | 25 µL |
| $ddH_2O$ | Up to 50 µL |

**2.** Run the following protocol in a thermocycler:

| **Primer extension reaction** | | |
|---|---|---|
| Step | Temperature | Time |
| Denaturation | 98 ˚C | 1 minute |
| Annealing* | 59 ˚C | 30 seconds |
| Extension | 72 ˚C | 30 seconds |
| Storage | 4 ˚C | Hold |

**to be adjusted depending on the primer used.*

**3.** Purify the double-stranded oligonucleotides using a PCR clean-up kit and elute with 25 µL elution buffer or sterile $ddH_2O$.

**3. Golden Gate reaction**. A Golden Gate reaction needs to be prepared that uses the double-stranded oligonucleotides to insert them into the destination vector containing the GFP-dropout cassette in the corresponding segment of your gene. Follow the steps:

**1.** Mix the following components in a PCR tube on ice:

| **Golden Gate reaction mix** | |
|---|---|
| Reagent | Volume |
| T4 ligase buffer 10x | 2.5 µL |
| T7 ligase | 1 µL |
| BsaI | 1 µL |
| Destination vector (20 ng/µL) | 1 µL |
| Double-stranded oligonucleotides (from previous reaction) | 20 µL |
| $ddH_2O$ | Up to 25 µL |

**2.** Run the following procedure in a thermocycler.

| **Golden Gate reaction** | | | |
|---|---|---|---|
| Step | Process | Temperature | Time |
| 1 | Digestion | 42 ˚C | 2 min |
| 2 | Ligation | 16 ˚C | 5 min |
| | Steps 1 – 2 (25x) | | |
| 3 | Final digestion | 60 ˚C | 10 min |
| 4 | Enzyme inactivation | 80 ˚C | 10 min |
| 5 | Storage | 4 ˚C | Hold |

**4. Transformation of *E. coli* with the library.** We use an in-house protocol for making *E. coli* DH5α cells chemically competent.[34] For the transformation, follow the steps:

**1.** Mix the entire 25 µL Golden Gate reaction with 150 µL of thawed competent cells in a 1.5 mL microcentrifuge tube.
**2.** Incubate this mixture on ice for 30 minutes.
**3.** Heat shock the cells for 45 seconds in a pre-heated water bath or thermomixer at 42°C.
**4.** Incubate the cells on ice for two minutes before adding 1 mL of LB medium.
**5.** Incubate the tube at 37 °C in a shaking or standing incubator for 1 hour.
**6.** Centrifuge the tube at 7,000 $g$ for 3 minutes to obtain a cell pellet.
**7.** Partially discard the supernatant to retain a volume of approximately 100 µL.
**8.** Resuspend the cell pellet by gently pipetting up and down. Plate the cell suspension on multiple LB agar plates with the corresponding antibiotic for plasmid maintenance (for our destination vectors 100 µg/mL ampicillin) using sterile glass beads or a cell spreader.
**9.** Incubate the plates overnight in a 37 °C stove.

**5. Evaluation of the library coverage and considerations for oversampling.** The design of the destination vectors allows for a simple Green-White fluorescent screening to assess whether the Golden Gate reaction was successful. The green fluorescent colonies still harbor the original plasmid (with the GFP-dropout cassette) while the white colonies carry the plasmids in which the Golden Gate reaction was very likely successful.

**1.** The green color from colonies with an unsuccessful Golden Gate reaction is visible to the naked eye after incubation of transformants for several hours in the fridge but it can also be visualized using a transilluminator or transilluminating goggles and a UV flashlight.
**2.** The efficiency of the Golden Gate reaction is calculated via **Equation 1** and requires counting the total amount of colonies and the number of green colonies. We routinely reach $10^3$ colonies per reaction and Golden Gate efficiencies of >97 % **(Supplementary Table 5** shows representative data from our library constructions). If the total colony number is lower than 100 or the Golden Gate efficiency is lower than 70 % see **Notes 3-5** for troubleshooting the Golden Gate reaction.

**Equation 1:**

$$Golden\ Gate\ efficiency\ (\%) = \left(1 - \frac{green\ colonies}{total\ colonies\ (green + white)}\right) x100$$

**3.** In order to ensure the presence of at least one representative of each possible variant in the library, estimate the expected library completeness, or vice versa calculating the number of white transformants required to achieve a high likelihood that all variants are represented. Use the tool GLUE for calculating library completeness.[35] GLUE uses the Poisson distribution to estimate the fractional completeness of a library, where $L$ is the number of white colonies, $V$ is the number of equiprobable variants. The probability that a variant occurs at least once is calculated via **Equation 2:**

$$1 - P(0) = 1 - e^{-L/V}$$

Note that this calculation only serves as an estimate of library coverage.
By simply using the number of white colonies for library size calculation, we likely overestimate the library coverage. This calculation relies on the assumption that every white colony contains a plasmid with a successful product from the Golden Gate reaction, and that all possible variants are equally distributed. However, part of the white colonies may contain undesired constructs due to synthesis errors that were present in some oligos. In addition, the distribution of colonies is likely not equimolar due to stochastic differences in the number of full-length molecules of each oligo and potential differences in cloning efficiency for different oligos. Nevertheless, this method still constitutes a good estimation procedure.
Example 1: For an insertional ALFA-tag scan of segment 2, 25 insertions are created (an insertion after each residue). To reach a likelihood of >0.95 that each variant is represented, ≥73 white colonies (*L*) need to be picked; to reach a likelihood of >0.98, ≥96 colonies need to be pooled **(Supplementary Figure 2A).**
Example 2: For the full *Venus* ORF 238 different variants are possible ($V = 238$).
To reach a likelihood of >0.95 that each variant is represented, ≥710 white colonies (*L*) need to be pooled; and to reach a likelihood of >0.98, ≥940 colonies need to be pooled **(Supplementary Figure 2B).**
**4.** Typically, the protocol delivers in the order of $10^2$-$10^3$ white colonies for a single round of Golden Gate assembly and transformation into *E. coli*. In case the required number of white colonies for high library coverage has not been reached after a single round of Golden Gate and transformation, one could perform more rounds in parallel.

**6. Pooling of the library.** The white *E. coli* colonies for each segment are counted and pooled segment-wise.
**1.** For each segment, prepare a sterile microcentrifuge tube with 1 mL LB media.
**2.** Use a sterile toothpick to scrape all white colonies, avoiding touching any green colonies. Dissolve the cells on the tip of the toothpick in the LB media.
**3.** Use 200 µL to inoculate a 2 mL overnight culture, specifically 2 mL LB media supplemented with the corresponding antibiotic (for our destination vectors 100 µg/mL ampicillin) and incubate for 16-20 hours at 37 °C in a shaking incubator.
**4.** From the remaining 800 µL, prepare a glycerol stock for future use by adding 200 µL 50 % (v/v) glycerol, transferring it to a cryo-tube and freezing it at -80 °C.
**5.** The next day, extract the plasmids from the overnight culture using a standard plasmid Miniprep kit.

**7. Transformation of yeast with the library.** For yeast transformation, we use a modified version of the protocol from Elble,[36] but likely other yeast transformationprotocols can be used. In terms of required yeast transformants (single colonies), oversample the number of variants in your library using the protocol provided in 3.1.2 step 5. We perform a separate transformation for each segment library, keeping a high level of control over the number of yeast colonies per segment, but mixing the segment libraries into one final library and performing a single transformation is also possible.
**1.** Prepare an overnight culture (1 mL per transformation) by inoculating *S. cerevisiae* BY4741 in YPD media and incubate for 16-20 hours at 30 °C with agitation.

**2.** Transfer 1 mL of the overnight culture into a microcentrifuge, pellet the cells at 8,000 *g* for 1 minute and decant the supernatant until ~50 µL of the liquid remains. Resuspend the cells gently.
**3.** Add 2 µL of 10 mg/mL single-stranded carrier DNA (*e.g.* from salmon sperm).
**4.** Add 1-2 µg of plasmid DNA (the library extracted in 3.1.2 step 6) and vortex briefly.
**5.** Add 500 µL of plate mixture and vortex until the cells and the viscous plate mixture have mixed.
**6.** Add 20 µL of 1 M DTT and vortex.
**7.** Incubate on your benchtop for 24 hours. During this time the cells will settle to the bottom of the tube.
**8.** Heat shock the cells for 10 min at 42 °C.
**9.** Pipette 75 µL from the settled cells and distribute them over four selective plates (in our case SC dropout media lacking histidine), spread with glass beads or a spatula and incubate the plate at 30 °C for two to three days.

**8. Considerations for selection and screening assays.** In order to evaluate the performance of each variant in the protein library, a relevant assay system is required that allows for sufficient throughput to assay a significant portion of the library. Having access to a suitable assay system that can be used to reach high library coverage is a crucial consideration before initiating the library creation.
**1.** The assay depends on the protein of interest and the research goal. Goals may range from identifying a limited number of variants showing improved protein activity to screening many variants to create a comprehensive sequence-fitness map: Appropriate assay systems are available for identifying new binders (*e.g.* yeast display), for enzyme function and enzyme kinetics and antimicrobial activity, among others.[2]
**2.** One generally distinguishes between selections and screens (which range from low to high throughput). In a selection, one uses conditional differences in cell growth, allowing one to enrich fast-growing cells from a pool of non-growing or poorly growing cells over time. Thus, a large library can be processed in one pot without individual evaluation of the performance of each variant. In a screen, every variant is evaluated individually by a dedicated assay (*e.g.* measurement of enzymatic activity or antimicrobial activity) and the screenable library size is therefore often much smaller than in a one-pot growth selection assay.
**3.** Depending on the selection or screen that will be used to assay the variant library, the yeast colonies obtained in step 2.7 can either be pooled (for growth selections) and frozen in a similar way to the *E. coli* library (see 3.1.2. step 6) or arrayed into 96- or 384-well plates (for medium throughput screens). In case of a screen, one needs to reach single colonies on a plate after transformation such that they can be individually picked and arrayed. In case colonies are too dense, they can be pooled, diluted 1:100 and 1:1000 and plated again.
**4.** Arraying is done by filling each well of a 96-well plate with 100 µL 2-times concentrated growth media using a multichannel pipette, followed by adding 50 µL sterile $H_2O$. Then, inoculate single colonies into each well using a sterile tooth picks, followed by incubation of the plate for 16-20 hours at 30 °C. Pick enough colonies to reach a high likelihood that every variant will be assayed (see 3.1.2 step 5). The arrayed plate can be directly used for screening or stored at -80 °C by adding 50 µL sterile 50 % (v/v) glycerol to each well.

**9. Considerations and resources for NGS and analysis.**

**1.** After the creation of the library, we recommend using NGS in order to determine its quality and completeness. This step can be performed directly after the plasmid preparation from *E. coli* or after transformation of the yeast, by pooling all the yeast colonies that are going to be assayed.

**2.** After screening or selection, the phenotype of the variants will need to be linked to their genotype. If only a few variants are selected, their genotype can be identified individually by Sanger sequencing. If the entire library has to be genotyped and linked to a phenotype, NGS approaches are more cost-efficient.

**3.** Several points need to be considered when generating amplicons for NGS:

**3.1.** NGS can be outsourced and there are different providers that offer well-suited services for variant library sequencing that allow small PCR amplified regions to be paired-end sequenced (~250 bases per forward and reverse read, pay attention to the maximal read length offered by each provider). If variant libraries are longer than the offered maximal read length, a tiling approach can be used, where contiguous sequences are amplified and covered by more than one probe.[37] Amplify the entire region of interest (the segment where the oligo pool was introduced) such that the coverage and sequencing depth are high. This purified PCR product is the product sent for NGS.

**3.2.** Partial Illumina adapters as well as barcodes for indexing can be incorporated into the amplicons during the PCR. In **Table 2,** we provide example primers (#8 and #9) that could be used for amplification of Segment 2. These primers encode for partial Illumina adaptors and generate a 260 bp amplicon (excluding the adaptors) that can be covered by a 250 bp forward and reverse read. Each read generates 250 bp, meaning the amplicon can be covered with a large overlap **(Supplementary Figure 1B).**

**3.3.** Use a limited amount of cycles (~20) and high-fidelity polymerase to limit the introduction of nucleotide substitutions during PCR for NGS sample preparation.

**3.4.** Use a fluorescence-based DNA quantification kit for accurate dsDNA quantification.

**3.5.** If variants from different samples, cultures, or bins are multiplexed, it is common practice to determine the DNA concentration of the amplicons before mixing them together in equimolar concentrations, in order to keep the reads per variant as similar as possible.

**3.6.** For analysis, nowadays no advanced bioinformatic experience is necessary; There are tools to analyze NGS sequencing data that enable inexperienced researchers to do such work.[37]

**4.** In case only a low or medium throughput assay is available, several tools for "*sequencing before screening*" have been developed that reduce the screening effort by reducing the required oversampling size of the library.[38][39] The tools use NGS before any screening effort has been made and rely on 96-well plate arrayed colonies and 96-well arrayed barcoded primers to amplify the region of interest in each variant. During analysis of the NGS reads, these double barcodes can link the variant sequences to the corresponding location in the plate. In this way, one can select variants of interest (*e.g.,* three replicates) for screening and, as such, reduce the oversampling size required. These "*sequencing before* *screening*" approaches require more investment into setting up the technique, but have the advantage that no information is lost, as opposed to *sequencing after screening*, which often requires binning the colonies based on their phenotype, but likely requires a smaller number of primers.

**3.2. Method 2. Cloning sgRNA libraries using oligo pools and Golden Gate cloning.**
Here we outline a method for using oligo pools to create a single gRNA (sgRNA) library for CRISPRx applications. Of note, gRNA libraries are usually much larger than the mutational libraries outlined in Method 1. For example, a typical genome-wide gRNA interference library for *S. cerevisiae* that targets each promoter of each open reading frame (>5000 ORFs) with at least 10 gRNAs comprises >50.000 independent variants.[40][41] Usually several (more than 10) independent Golden Gate reactions and transformations need to be performed to reach the necessary library size.

To follow section 3.2. (Method 2), we structured it in analogy to section 3.1. (Method 1, **Figure 1)**. Several sub-steps of section 3.2.2. are identical to section 3.1.2. and we then refer to the respective sub-steps in section 3.1.2.

**3.2.1. Design of the library**: this includes 1. The design of the destination vector and 2. the design of the oligo pool.

**3.2.2**. **Creation of the library**: this includes 1. cloning of the destination vector; 2. double-stranding the oligos; 3. pre-digesting the destination vector; 4. performing the Golden Gate reaction; 5. transforming *E. coli* with the mutational library; 6. evaluating the library coverage and consideration for oversampling; 7. pooling the library; 8. transforming the library into yeast (*S. cerevisiae*);

Steps 5. to 8. are identical to steps 4. to 7. in section 3.1.2.

Further, we provide suggestions for library quality control via NGS (**9.**). This point strongly overlaps with step 8. in section 3.1.2.

**3.2.1. Design of the sgRNA library.**

**1. Design of the destination vector.**

We use the gRNA expression cassette featured in the YTK part pYTK50.[25] This expression cassette uses the well-established CRISPRm sgRNA architecture,[42] where the gRNA expression is driven by a phenylalanine tRNA, followed by an HDV self-cleaving ribozyme that is linked to the sgRNA to stabilize it in the cell; the expression is terminated by an SNR52 terminator **(Figure 5A).** For creating a Golden Gate compatible gRNA cloning system, the 20 bp targeting sequence was replaced by a BsmBI-flanked GFP-dropout cassette. pYTK50 is designed as an entry-level vector featuring only replication and selection components for *E. coli.* In order to create a Golden Gate compatible destination vector that yields an expression-ready sgRNA library in yeast, we cloned the pYTK50 sgRNA expression cassette into a pRS414-type vector that was originally designed for *Candida glabrata.*[27] In this way, the oligo pool library can be encoded as a 20 bp spacer region plus the compatible 4 bp overhangs and the BsmBI restriction site **(Figure 5).** Note that here we only provide a general protocol for designing compatible oligo pools. This protocol does not include the design of the sequences of the spacer regions, as those depend on the application and the specific yeast genome sequence. For that step, several examples and resources have been described.[40][41][43]

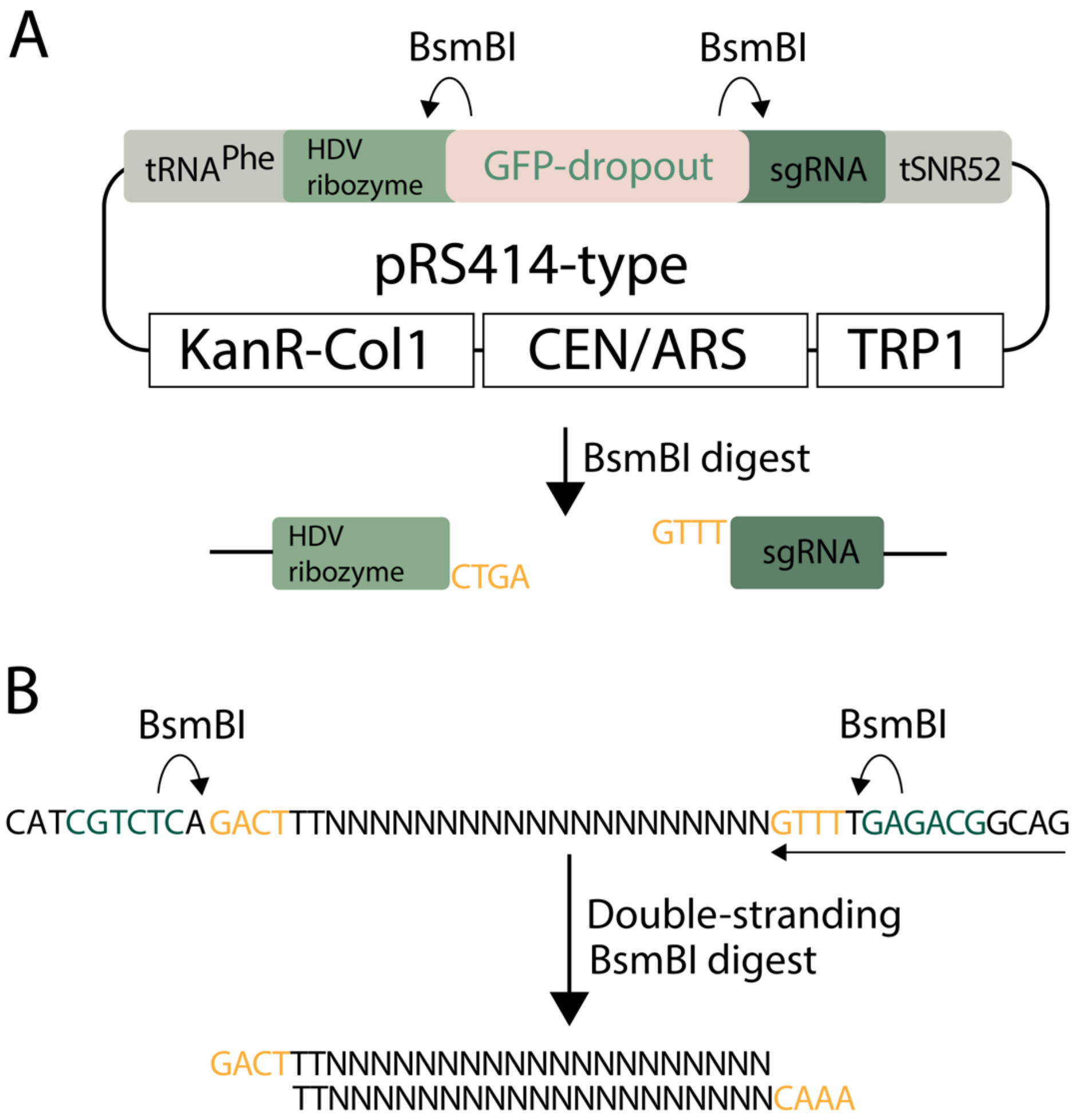

**Figure 6. Design of the gRNA destination vector and the spacer-encoding oligo pool. A.** The gRNA destination vector features an expression cassette based on a tRNA promoter, a self-cleaving HDV ribozyme followed by a GFP-dropout cassette that is replaced by the actual 20 bp targeting spacer during the Golden Gate reaction to yield a functional sgRNA and an SNR52 terminator. **B.** Each oligo is designed by using a distinct 20 bp targeting sequencing (here depicted as $N_x$), by adding the appropriate 4 bp overhangs at the 5' and 3' ends (yellow), followed by adding a 3' facing BsmBI recognition sites (green) and several nucleotides to enhance BsmBI-cutting. The bases also yield the reverse priming site (black arrow). After double stranding, the oligos are added to the Golden Gate reaction where the BsmBI digestions yield the correct overhangs for cloning into the destination vector.

**2. Design of the spacer-encoding oligo pool.**

Here we outline the general design of the spacer-encoding oligo pool. Follow the steps **(Figure 5):**

**1.** Generate a list of 20 bp spacer sequences based on your intended CRISPR application and target organism; examples and resources have been described.[40][41][43]

**2.** Add the four-base overhangs GACT and GTTT to the 3' and 5' ends that allow for cloning into the destination vector.

**3.** Add two thymines (Ts) between the overhangs and the 20 bp spacer. This is as the 5' sequence gRNA design has two thymine bases flanking the targeting region.[25] As this would lead to a 5' CTTT overhang, which can potentially mis-ligate with the 3'GTTT overhang, the two thymines are incorporated in the oligo.

**4.** Add 3' cutting BsmBI-recognition sites to the 5' and 3' ends of each oligo.

**5.** Add at least three extra nucleotides to improve the digestion efficiency. Here we add CAT at the 5' and GCAG at the 3' end, also yielding a primer landing pad that can be used for double-stranding the oligos from the library **(Figure 5B and Table 2)**. The resulting sequence of this double-stranding reverse primer is 5' – CTGCCGTCTCAAAAC – 3'.

**3.2.2. Creation of the sgRNA library.**

**1. Cloning of the destination vector.**

In analogy to section 3.2.1, we provide one example of how to clone the destination vectors. However, these can be cloned using any molecular cloning method of choice.

We assembled the destination vector as follows: The pRS414-type vector was assembled via Golden Gate using the parts pYTK002, pYTK047, pYTK072, pCgTK02, pCgTK01, and pYTK084 **(Table 1).** The gRNA expression cassette was amplified by PCR using pYTK50 as a template and primers 11 and 12 (**Table 2**) to add BsaI overhangs. The pRS414-type expression vector was digested with BsmBI, and the gRNA expression cassette-encoding PCR product was cloned in via a BsaI-based Golden Gate reaction.

**2. Making double-stranded oligos.**

In analogy to Method 1, we run a single primer extension reaction using the reverse primer that binds to the landing pad (sgRNA reverse primer: Table 2). Follow the steps:

**1.** Mix the following components in a PCR tube on ice:

| **Primer extension mix** | |
|---|---|
| Reagent | Volume |
| Single-stranded oligo pool (30 μM) | 1 μL |
| Primer (10 μM) | 2.5 μL |
| 2x Phire Hot Start II PCR Master Mix | 12.5 μL |
| Sterile water | Up to 25 μL |

**2.** Run the following protocol in a thermocycler:

| **Primer extension reaction** | | |
|---|---|---|
| Step | Temperature | Time |
| Denaturation | 98 ˚C | 1 minute |
| Annealing* | 55 ˚C | 30 seconds |
| Extension | 72 ˚C | 1 minute **(Note 7)** |
| Store | 4 ˚C | Hold |

** Adjust temperature based on the used primer*

**3.** Purify the double-stranded oligonucleotides obtained using a PCR clean-up kit and elute with 25 µL sterile water.

### 3. Pre-digestion of the destination vector

We observed that digesting and gel-purifying the destination vector prior to the Golden Gate reaction greatly increased the Golden Gate efficiency for this protocol **(Note 8).** Follow the steps:

**1.** For digesting the vector mix the following components in a microcentrifuge tube on ice.

| **Vector digest mix** | |
|---|---|
| Reagent | Volume |
| NEB Buffer 3.1 10x | 5 µL |
| Destination vector (100 ng/µL) | 10 µL |
| BsmbI-v2 | 1 µL |
| Sterile water | Up to 50 µL |

**2.** Incubate for 2 hours at 55 ˚C.

**3.** Separate the digested vector by using standard agarose gel electrophoresis. Cut the band that corresponds to the destination vector with the GFP-dropout cassette being released and purify it using a standard Gel purification kit.

**4.** Measure the concentration of the Gel-purified digested destination vector using a spectrophotometer.

### 4. Golden Gate reaction

The Golden Gate reaction needs to be prepared using the double-stranded oligonucleotides (insert) and the digested destination vector. Follow the steps:

**1.** Mix the following components in a PCR tube on ice:

| **Golden Gate reaction mix** | |
|---|---|
| Reagent | Volume |
| T4 ligase buffer 10x | 2.5 µl |
| T7 ligase | 1 µl |
| BsmbI-vs2 | 1 µl |
| Digested destination vector (10 ng/µl) | 1 µl |
| Double-stranded oligonucleotides (50 ng/µl) | 1 µl |
| Sterile water | Up to 25 µl |

**2.** Run the following procedure in a thermocycler. Note that this protocol uses 60 cycles of digestion and ligation instead of the 25 cycles outlined in section 3.1.2 step 3. We found that increasing the cycle number increased the Golden Gate efficiency for this destination vector.

| **Golden Gate reaction** | | | |
|---|---|---|---|
| Step | Process | Temperature | Time |
| 1 | Digestion | 42 ˚C | 2 min |
| 2 | Ligation | 16 ˚C | 5 min |
| | Steps 1 – 2 (60x) | | |
| 3 | Final digestion | 60 ˚C | 10 min |
| 4 | Enzyme inactivation | 80 ˚C | 10 min |
| 5 | Storage | 4 ˚C | Hold |

**5. Transform *E. coli* with the library.**
Perform as outlined in **3.1.2.** step 4

**6. Evaluation of the library coverage and consideration for oversampling.**
Perform as outlined in **3.1.2.** step 5. Be aware that sgRNA libraries are usually much larger than mutational libraries.

**7. Pooling of the library**
Perform as outlined in **3.1.2.** step 6.

**8. Transformation of *S. cerevisiae* or *C. glabrata* with the library.**
For *S. cerevisiae* perform as outlined in **3.1.2.** step 7. For *C. glabrata* use the transformation protocol described by Gietz *et al.*.[44]

**9. Suggestions for library quality via NGS.**
The recommendations are similar to those in **3.1.2.** step 8.
A few colonies can be picked and Sanger sequenced. The completeness and quality of the library should then be checked by NGS. The 20 bp spacer region can be amplified by PCR using primers 13 and 14 suggested in **Table 1,** generating a ~250 bp amplicon. The primers already contain partial Illumina adapters that can be used as templates for i7 and i5 index primers (see **Note 6**). Single-end short-read sequencing (100 cycles) should be sufficient to capture the 20 bp spacer sequences within the amplicons.

## 4. Notes

**1.** The price of the oligo pool depends mainly on the total number of ordered nucleotides and vendors often specify the price per nucleotide. Two examples on how to calculate the number of nucleotides is given in **Supplementary Note 1 and 2**.

**2.** Primers #1 and #2 amplify the 5' region flanking segment 2, primers #3 and #4 amplify the 3' region flanking segment 2, and primers #5 and #6 amplify the GFP dropout cassette (see **Supplementary Figure 1 and Table 2**).
**3.** The transformation of a successful Golden Gate reaction should yield $10^2$ to $10^3$ colonies per reaction with an efficiency of at least 95%.
**4.** If few colonies appear on selective plates after transformation of the Golden Gate reaction, verify the following: (i) competence of the cells used. We recommend testing competence with a purified plasmid or designing an easy "control Golden Gate reaction" based on the well-established YTK. (ii) the functionality of the T7 ligase: zero or a low number of colonies can be caused by poor ligation efficiency regardless of the digestion. If the destination vector is digested but not ligated during the Golden Gate reaction, the competent cells used will not be transformed, resulting in no or low number of colonies. Perform a "control Golden Gate reaction" as mentioned above or order fresh T7 ligase.
**5.** If a high number of green fluorescent colonies appear on selective plates after transformation of the Golden Gate reaction, do the following: (i) check functionality of the BsaI or BsmBI enzymes by perform a control digest, or order fresh BsaI / BsmBI. Transformation of undigested destination vectors will lead to the growth of cells with intact plasmids containing the GFP-dropout cassette. (ii) Pre-digest the destination vector and gel-purify it before adding it to the Golden Gate reaction as described in section 3.2.2 step 3.
**(iii)** Increase the number of cycles from 25 to 60 during the Golden Gate reaction.
The above tips were inspired by our own experience and resources on the BarrickLab website.[45]
**6.** To minimize PCR bias during NGS sample preparation, it is recommended to run a qPCR on isolated library material (the prepped plasmid library) to determine a low PCR cycle number that is still within the dynamic range of detection. If too many PCR cycles are chosen each guide will be amplified to saturation and no difference in abundance between the guides can be determined.
**Note 7:** This double-stranding protocol was established with 1-minute extension time. This differs from 30-seconds extension time of Method 1, but we have not specifically optimized the extension time in either protocol. Both extension times worked well for us.
**Note 8:** This step is mainly recommended when one needs a high number of (white) colonies

## Funding

This work was partly funded by NWO grant OCENW.M20.250 (SB).

## Author contributions

AMV, TdV, RCP and SB developed the method. AMV, TdV, and RCP drafted the method description. SB wrote the manuscript. All authors edited and approved the manuscript.

# Supplementary Material

**Combining oligo pools and Golden Gate cloning to create protein variant libraries or guide RNA libraries for CRISPR applications.**

Maciá Valero *et al.*

## Supplementary Figures

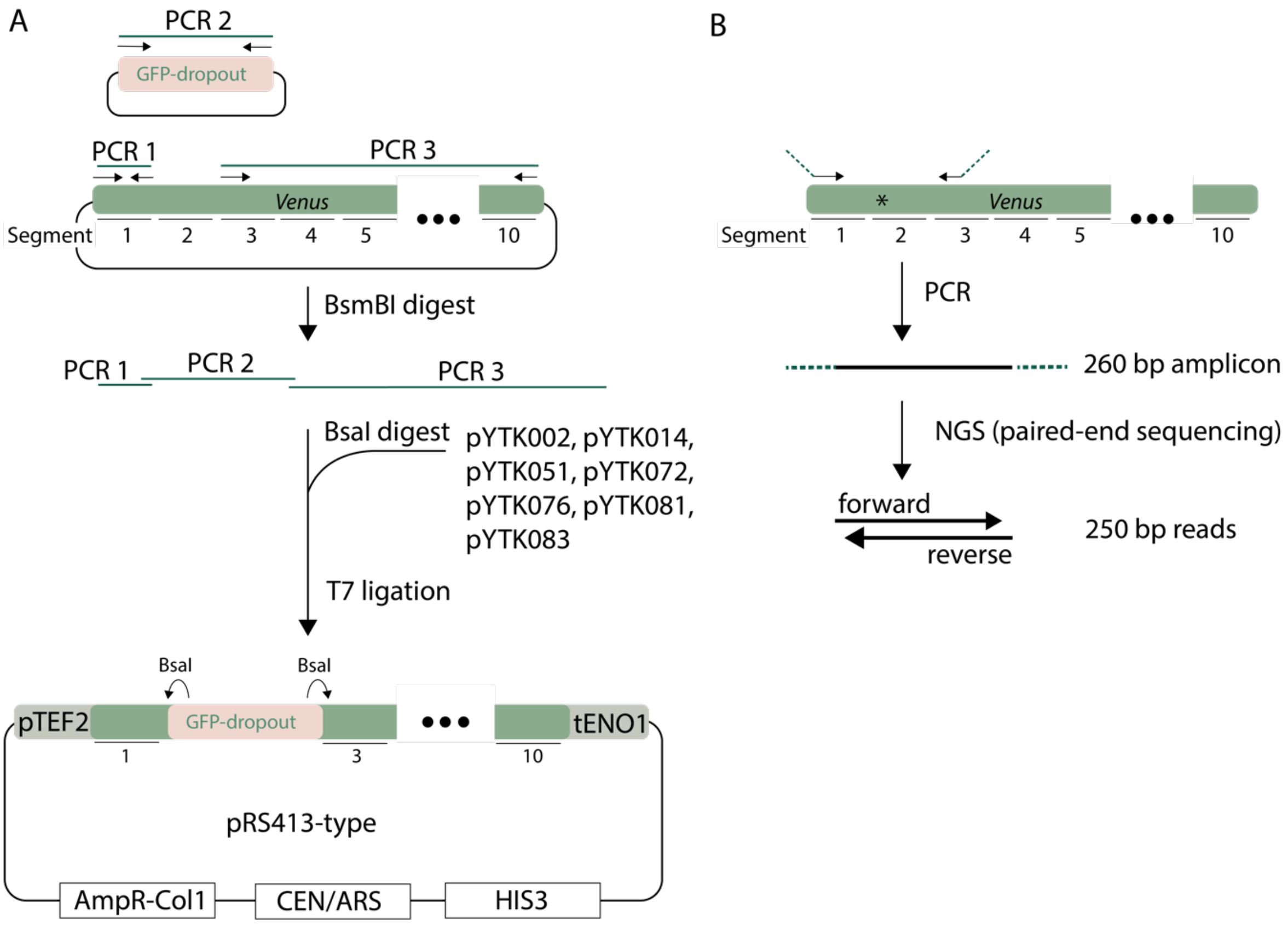

**Supplementary Figure 1. Suggestion for constructing a Destination vector and amplifying regions of a protein for paired-end NGS analysis. A.** For cloning Destination vector 2, the regions flaking segment to at the 5' (PCR1) and 3' side (PCR3) as well as the GFPdropout cassette (PCR2) should be amplified with primers 1-6 **(Table 2).** The primers encode for BsmBI restriction sites. The PCR products should then be digested with BsmBI to generate the appropriate overhangs. Further, the indicated YTK parts should be BsaI digested. Both enzymes should be heat-inactivated and all components can be mixed and ligated. **B**. For paired-end NGS, individual segments can be amplified by PCR. Primers should be chosen based on the sequencing length offered by a provider. In our case we use a paired-end amplicon sequencing service yielding a forward and a reverse read of 250 bp. Table 2 provides two example primers that can be used to amplify a 260 bp fragment containing Segment 2. The overlapping primers allow for an additional sequencing quality control, given each molecule is sequenced twice.

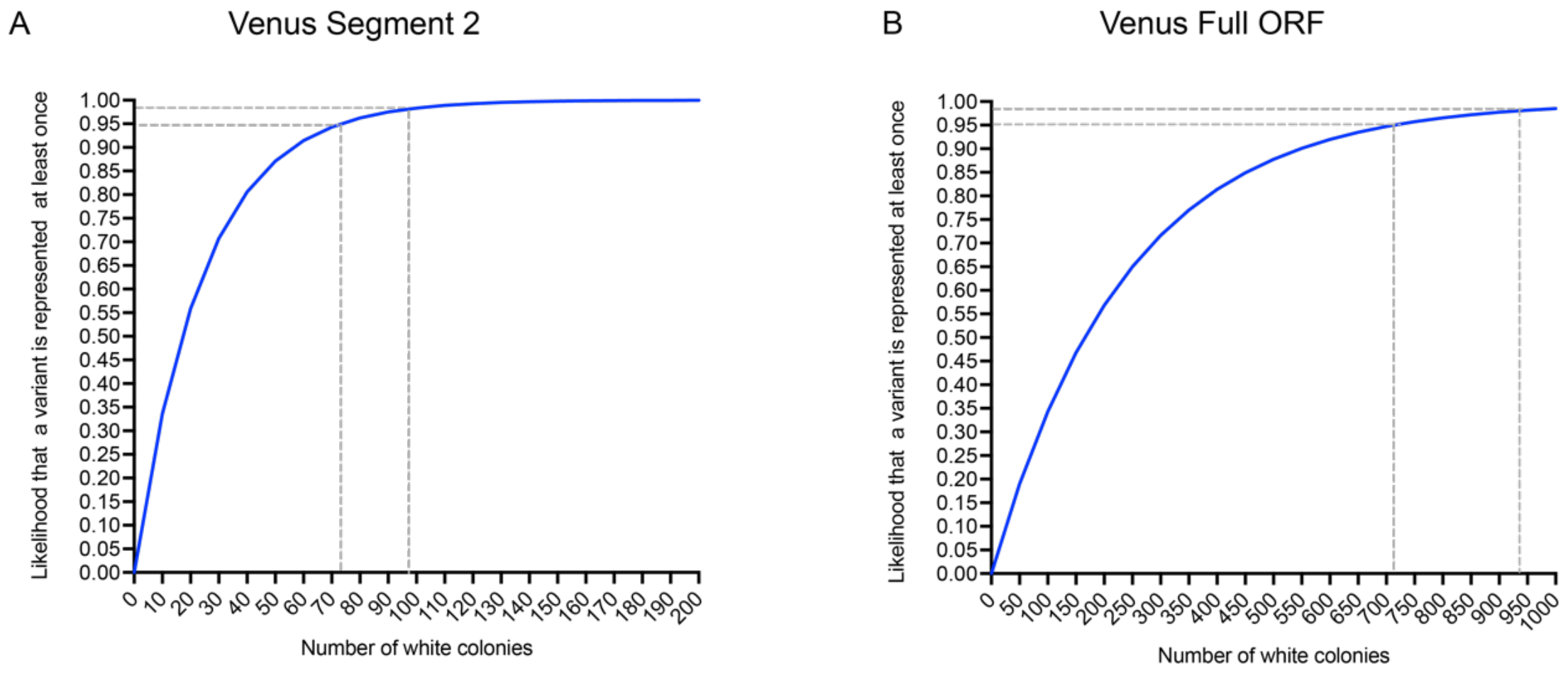

**Supplementary Figure 2. Library Coverage.** Example of calculations of library coverage using GLUE.[1] The equation $1 - P(0) = 1 - e^{(-L/V)}$ was plotted against an increasing number of white colonies (L). **A.** For the Venus protein Segment 2 insertion scanning library, 25 equiprobable variants (V) are expected. The number of white colonies needed to be pooled in order to reach a likelihood of 0.95 and 0.98 that all variants are present is at least 73 and 96, respectively. **B.** For scanning the full *Venus* ORF, 238 equiprobable variants (V) are expected. To reach a likelihood of 0.95 that all variants are present in the library at least once, more than 710 white colonies need to be pooled; to reach a likelihood of 0.98, at least 940 colonies need to be pooled.

## Supplementary Tables

**Supplementary Table 1.** Sequences of the *in silico* segmented Venus gene within pRS413-pTEF2-Venus-tENO1 and the 10 destination vectors.

| **Name** | ***Benchling* link** |
|---|---|
| pRS413-TEF2-Venus-ENO1 | https://benchling.com/s/seq-oJbVApv9HJmQzIGW6Jpd?m=slm-F2O1rfg27crWoO9z1BVH |
| Destination vector 1 | https://benchling.com/s/seq-WUMU0NpDakx6DoLg5TuV?m=slm-1NVA24UFx64BOGSGXO3F |
| Destination vector 2 | https://benchling.com/s/seq-oKpIgHPbRI4lN3JW7gow?m=slm-T6NwsAe3Pvw8wRsi1qmm |
| Destination vector 3 | https://benchling.com/s/seq-2lvPfYP6JvWkqHLTid5i?m=slm-UpO5UJCbYS2qisUgUP4A |
| Destination vector 4 | https://benchling.com/s/seq-K52SpAqBQSy8ROBeWdUr?m=slm-xLeUswP2WRM4dlnXSFvr |
| Destination vector 5 | https://benchling.com/s/seq-nV6lUGEZOSgwrckAWqBY?m=slm-Hct1C7Ox6r6SL5SHE5FK |
| Destination vector 6 | https://benchling.com/s/seq-iqVhyVY8xIpBFSG65NkQ?m=slm-GO5lyaKOXnXqPLySONTj |
| Destination vector 7 | https://benchling.com/s/seq-856aDgRTA5t0wqwnrs1i?m=slm-CSIridihVnjmvwqraA38 |
| Destination vector 8 | https://benchling.com/s/seq-u8EVXsk9C5pgzsQ2Da4D?m=slm-xLOUflV1zEi7OTK8HlIO |
| Destination vector 9 | https://benchling.com/s/seq-YoZf2HFHnuFa1srJy8el?m=slm-bSAAgo327IsgOnjT54h8 |
| Destination vector 10 | https://benchling.com/s/seq-3ZtpRIoPTmhvep2BfemX?m=slm-eGDt7yqe3QunsP3Q5bYf |
| gRNA destination vector | https://benchling.com/s/seq-aECx9DtiS8oa99gQifBm?m=slm-baqd6hS7wXlNRZmQohyo |

**Supplementary Tables 2-4 can be requested from the authors.**

**Supplementary Table 5. Golden Gate efficiency.** Data from an in-house library assembly are presented, including the total number of colonies after transformation of the Golden Gate reaction mixes into *E. coli*, the number of green colonies on the plates, and the calculated Golden Gate efficiency based on these numbers. The efficiency can depend on the chosen overhangs, which are given in the first column. **Note:** The Golden Gate efficiency calculated here only represents part of the total efficiency. The total number of colonies should be at least $>10^2$. If very few colonies are found than likely the ligation efficiency was very low.

| Overhangs 5' and 3' | Total number of colonies (rounded) | Green colonies | Golden Gate efficiency |
|---|---|---|---|
| gcgt and ggta | 1200 | 0 | 100 % |
| cacg and ggag | 1000 | 13 | 98.7 % |
| cata and aagc | 1000 | 23 | 97.7 % |
| tggg and agat | 1000 | 3 | 99.7 % |
| gggt and aaca | 800 | 15 | 98.2 % |
| catc and atcc | 1100 | 0 | 100 % |

## Supplementary Notes

### Supplementary Note 1. Calculation of oligo pool size.

Oligo pool price depends on the number of total bases, which can be calculated as follows:

Example 1. Insertional scanning: In our case, each segment has a different size, and accordingly a different number of bases. Each segment contains 39 nt (ALFA-tag) + 8 nt (overhangs) + 14 nt (BsaI sites) + 20 (extra nts) = 81 nt. For each segment, add the number of nucleotides and multiply the result by the number of codons (positions where the tag will be inserted) that are present.

- Segment 1 → (81 nt + 72 nt) x 24 = 3,672 nt
- Segment 2 → (81 nt + 75 nt) x 25 = 3,900 nt
- Segment 3 → (81 nt + 72 nt) x 24 = 3,672 nt
- Segment 4 → (81 nt + 81 nt) x 27 = 4,374 nt
- Segment 5 → (81 nt + 69 nt) x 23 = 3,450 nt
- Segment 6 → (81 nt + 75 nt) x 24 = 3,900 nt
- Segment 7 → (81 nt + 75 nt) x 25 = 3,900 nt
- Segment 8 → (81 nt + 75 nt) x 25 = 3,900 nt
- Segment 9 → (81 nt + 63 nt) x 21 = 3,087 nt
- Segment 10 → (81 nt + 60 nt) x 19*[1] = 2,679 nt

3,672 nt (seg 1) + 3,900 nt (seg 2) + 3,672 nt (seg 3) + 4,374 nt (seg 4) + 3,450 nt (seg 5) +

3,900 nt (seg 6) + 3,900 nt (seg 7) + 3,900 nt (seg 8) + 3,087 nt (seg 9) + 2,679 nt (seg 10) = 36,534 nucleotides.

*[1]Note: Segment 10 includes the stop codon. No insertion is introduced at this position; therefore, the number of nucleotides should be multiplied only by translated codons. The *start* and *stop* codon are included in the library given as an example but, as none should be modified during the cloning process, both can be excluded from the library, which would reduce slightly the cost of the library quotation.

<u>Example 2. Mutational scanning</u>: Each oligo contains 8 nt (overhangs) + 14 nt (BsaI sites) + 20 (extra nts) = 42 nt. Subtract the number of positions for which certain conversions would result in the wildtype sequence, which are excluded from the oligo pool.

- Seg 1 → (42 nt + 72 nt) x 23*[2] codons x 7 mutations - 8 x (42 nt + 72 nt) = 17,442 nt
- Seg 2 → (42 nt + 75 nt) x 25 codons x 7 mutations - 8 x (42 nt + 75 nt) = 19,539 nt
- Seg 3 → (42 nt + 72 nt) x 24 codons x 7 mutations - 10 x (42 nt + 72 nt) = 18,012 nt
- Seg 4 → (42 nt + 81 nt) x 27 codons x 7 mutations - 11 x (42 nt + 81 nt) = 21,894 nt
- Seg 5 → (42 nt + 69 nt) x 23 codons x 7 mutations - 8 x (42 nt + 69 nt) = 16,983 nt
- Seg 6 → (42 nt + 75 nt) x 25 codons x 7 mutations - 6 x (42 nt + 75 nt) = 19,773 nt
- Seg 7 → (42 nt + 75 nt) x 25 codons x 7 mutations - 6 x (42 nt + 75 nt) = 19,773 nt
- Seg 8 → (42 nt + 75 nt) x 25 codons x 7 mutations - 10 x (42 nt + 75 nt) = 19,305 nt
- Seg 9 → (42 nt + 63 nt) x 21 codons x 7 mutations - 10 x (42 nt + 63 nt) = 14,385 nt
- Seg 10 → (42 nt + 60 nt) x 19*[2] codons x 7 mutations - 7 x (42 nt + 60 nt) = 12,852 nt

17,442 nt (seg 1) + 19,539 nt (seg 2) + 18,012 nt (seg 3) + 21,894 nt (seg 4) + 16,983 nt (seg 5) + 19,773 nt (seg 6) + 19,773 nt (seg 7) + 19,305 nt (seg 8) + 14,385 nt (seg 9) + 12,852 nt (seg 10) = 179,958 nucleotides.

*[2]Note Segments 1 and 10 include the start and stop codon. No mutation is tested in those positions; therefore, the number of nucleotides should be multiplied only by mutated codons. These two codons are included in the example but, given that none should be modified during the cloning process, both can be excluded from the library, which would reduce slightly the cost of the order for synthesized oligos.